\documentclass{ws-ijmpe}

\begin{document}

\markboth{G.L.P. Silva, M.J. Menon, R.F. \'Avila}{Proton Profile Function
at 52.8 GeV}

\catchline{}{}{}{}{}

\title{PROTON PROFILE FUNCTION AT 52.8 GeV}

\author{\footnotesize GEOVANNA LUIZ PEREIRA DA SILVA 
and M\'ARCIO JOS\'E MENON}

\address{Instituto de F\'{\i}sica Gleb Wataghin, Universidade 
Estadual de Campinas \\
Campinas, SP 13083-970,Brazil \\
geovape@ifi.unicamp.br, menon@ifi.unicamp.br}

\author{REGINA FONSECA \'AVILA}

\address{Instituto de Matem\'atica, Estat\'{\i}stica e Computa\c c\~ao 
Cient\'{\i}fica \\
Universidade Estadual de Campinas \\
Campinas, SP 13083-970, Brazil \\
rfa@ifi.unicamp.br}

\maketitle

\begin{history}
\received{(10 May 2007)}
\accepted{(6 June 2007)}
\end{history}

\begin{abstract}
We present the results of a novel
model-independent fit to elastic proton-proton 
differential cross section data 
at $\sqrt s$ = 52.8 GeV.
Taking into account the 
error propagation from the fit parameters, we determine 
the scattering amplitude in 
the impact parameter space (the proton profile function)
and its statistical uncertainty region. 
We show that both the real and imaginary parts of the profile
are consistent with two dynamical contributions, one from a central
dense region, up to roughly 1 fm and another from a peripheral evanescent
region from 1 to 3 fm.
\end{abstract}

\section{Introduction}

In the absence of a pure QCD description of high-energy
\textit{soft diffractive processes},
elastic hadron scattering constitutes one of the topical
unsolved problems in Particle Physics.\cite{pred} At this stage, 
empirical information, extracted from the experimental data,
play a fundamental role in the construction
of phenomenological models and to establish connections
between experimental data
and first principles and theorems in the underlying quantum field theory.
With this strategy, we have investigated proton-proton ($pp$)
and anti\-proton\--proton ($\bar{p}p$) elastic scattering by means of model
independent approaches (\textit{the inverse problem}) and have extracted
several properties of these processes in the impact parameter
space.\cite{prev}
In these works we made use of an analytical parametrization for the
scattering amplitude as a sum of exponentials, but with constrained real 
and imaginary parts.
In this communication, we make use of a novel analytical parametrization 
for the amplitude without the above constraint and we also develop a detailed data 
reduction procedure which allows a better fit result on statistical
grounds.\cite{ms} We treat here only $pp$ scattering at
$\sqrt s$ = 52.8 GeV together with data
in the large momentum transfer region from $pp$ scattering
at 27.5 GeV. From this data reduction, we extract the complex Profile Function
(impact parameter space) and show that both the real 
and imaginary parts of the 
profile indicate two distinct contributions, one central and
one peripheral.

\section{Parametrization and Fit Results}

We consider here the following physical quantities that characterize the high-energy elastic
hadron scattering:\cite{pred} the differential cross section,

\begin{equation}
\frac{d\sigma}{dq^2}(s, q^2) = \pi|F(s, q^2)|^2,
\end{equation}
the total cross section (Optical Theorem) and the $\rho$ parameter,

\begin{equation}
\sigma_{\mathrm{tot}}(s) = 4\pi \textrm{Im}\ F(s, q^2=0),
\qquad
\rho(s) = \frac{\textrm{Re}\ F(s, q^2=0)}{\textrm{Im}\ F(s, q^2=0)},
\end{equation}
where $F$ is the scattering amplitude, $s$ and $q^2 = -t$ are the Mandelstam
variables.

The largest set of data presently available on differential
cross sections, as a function of the momentum transfer, concerns 
proton-proton scattering at $\sqrt s$ = 52.8 GeV.
The data
was \textit{carefully compiled and normalized} by Amaldi
and Schubert\cite{as1} and include the optical point,
$d\sigma/dq^2(q^2=0) = \sigma_{\mathrm{tot}}^2 (1 + \rho^2)/16\pi$,
with the average experimental values  $\sigma_{\mathrm{tot}}$ =
42.67 $\pm$ 0.19 mb
and $\rho$ = 0.078 $\pm$ 0.010.
Based on the independence of the experimental data with the
energy at large momentum transfer (ISR region),\cite{prev}
we also include in the analysis the data from $pp$ scattering at $\sqrt s$
= 27.5 GeV in the region $5.5 \leq q^2 \leq 14$ GeV$^2$.\cite{faissler}

We consider the parametrization for the
amplitude as a sum of exponentials in $q^2$, with both
real and imaginary parts connected with the forward experimental
data on $\sigma_{\mathrm{tot}}$ and $\rho$ through Eq. (2).
Denoting $a_{i}$, $b_{i}$, $i=1,2,...,m$ and $c_{j}$, 
$d_{j}$, $j=1,2,...,n$ 
the real free parameters associated with the real and imaginary
parts of the amplitude, respectively, the parametrization can be 
expressed by \cite{ms}

\begin{eqnarray}
F(s,q) &=& \left\{ \left[ {\rho \ \sigma_{\mathrm{tot}} \over 4 \pi} -
\sum_{i=2}^{m} a_i \right] e^{- b_1 q^{2}} +
\sum_{i=2}^{m} a_i e^{- b_i q^{2}} \right\} \nonumber \\
&+&
\mathrm{i} \left\{
\left[ {\sigma_{\mathrm{tot}} \over 4 \pi} -
\sum_{j=2}^{n} c_j \right] e^{- d_1 q^{2}} +
\sum_{j=2}^{n} c_j e^{- d_j q^{2}} \right\},  
\end{eqnarray}
allowing the elimination of two 
free parameters, which we choose above to be $a_1$ (real part) and $c_1$
(imaginary part). With this we have $2(m + n) - 2$ free
fit parameters, with $\sigma_{\mathrm{tot}}$ and $\rho$ given by the
corresponding experimental values quoted above.

With parametrization (3) we fit the $pp$ differential cross
section data at 52.8 GeV through Eq. (1) and the CERN-Minuit code.
Since we have no information on the contribution from the real part
of the amplitude beyond the forward region,
we started the fit by looking for the best result with all possible choices
of contributions, namely, ($m$, $n$) = (0, 1), (1, 1), (1, 2), (2, 1), (2, 2), 
etc..., until reaching the smallest $\chi^2$/DOF.
In a second step we added the data from 27.5 GeV,
testing in the same way different contributions from the real and imaginary parts.

In Fig. 1 we display the fit results to $pp$ data at $\sqrt s =$ 52.8 GeV
($q_{max}^2 =$ 9.8 GeV$^2$, left panel) and  
adding data at 27.5 GeV ($q_{max}^2 =$ 14 GeV$^2$, right panel), together with the corresponding experimental data 
and uncertainty
regions obtained by error propagation from the fit parameters.
In the first case we obtained $\chi^2$/DOF = 1.43 and in the second
$\chi^2$/DOF = 1.55, which are below our previous results,
$\chi^2$/DOF = 1.65 and  $\chi^2$/DOF = 2.07, respectively.\cite{prev}

\begin{figure}[th]
\centerline{\psfig{file=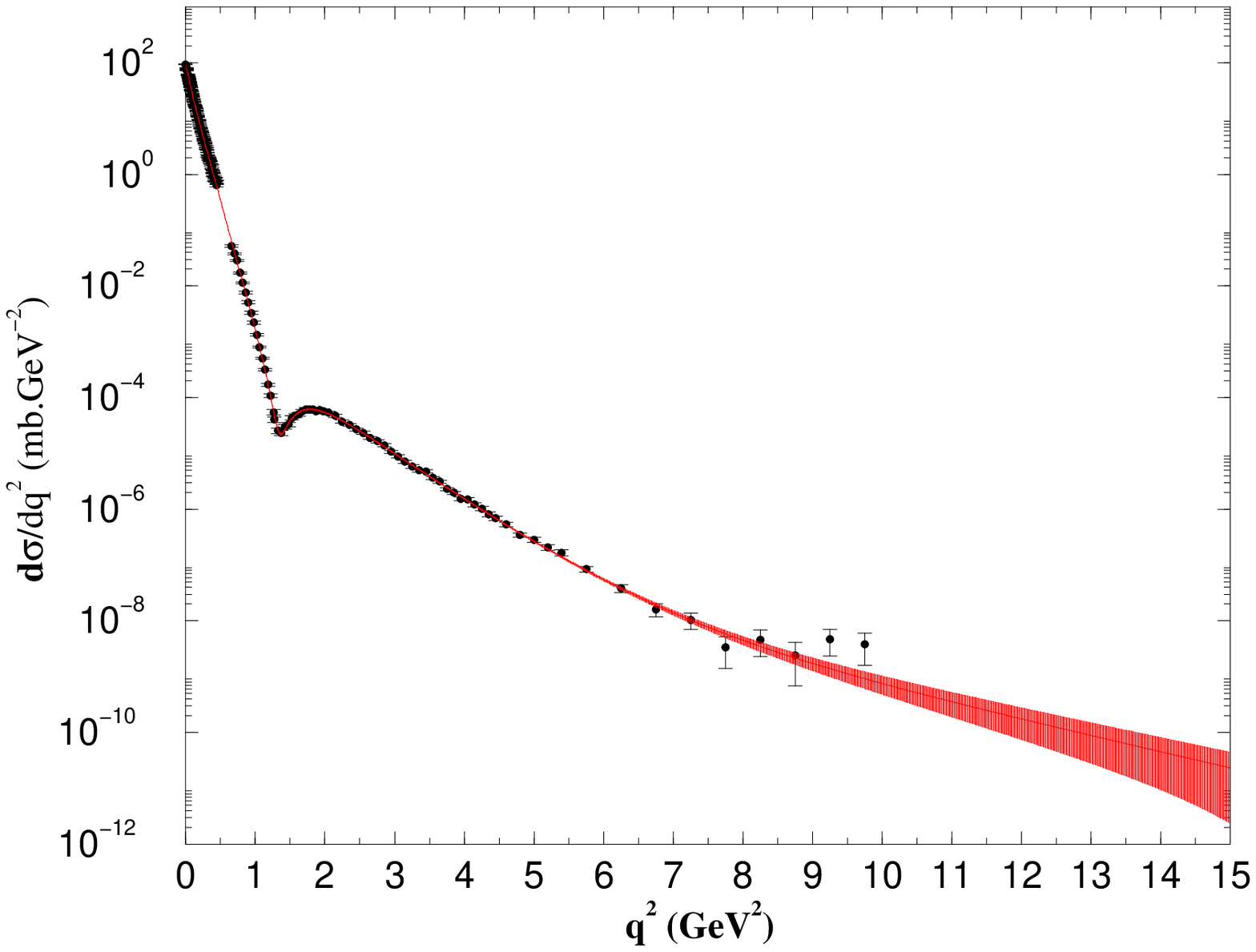,width=6cm,height=6.5cm}
\psfig{file=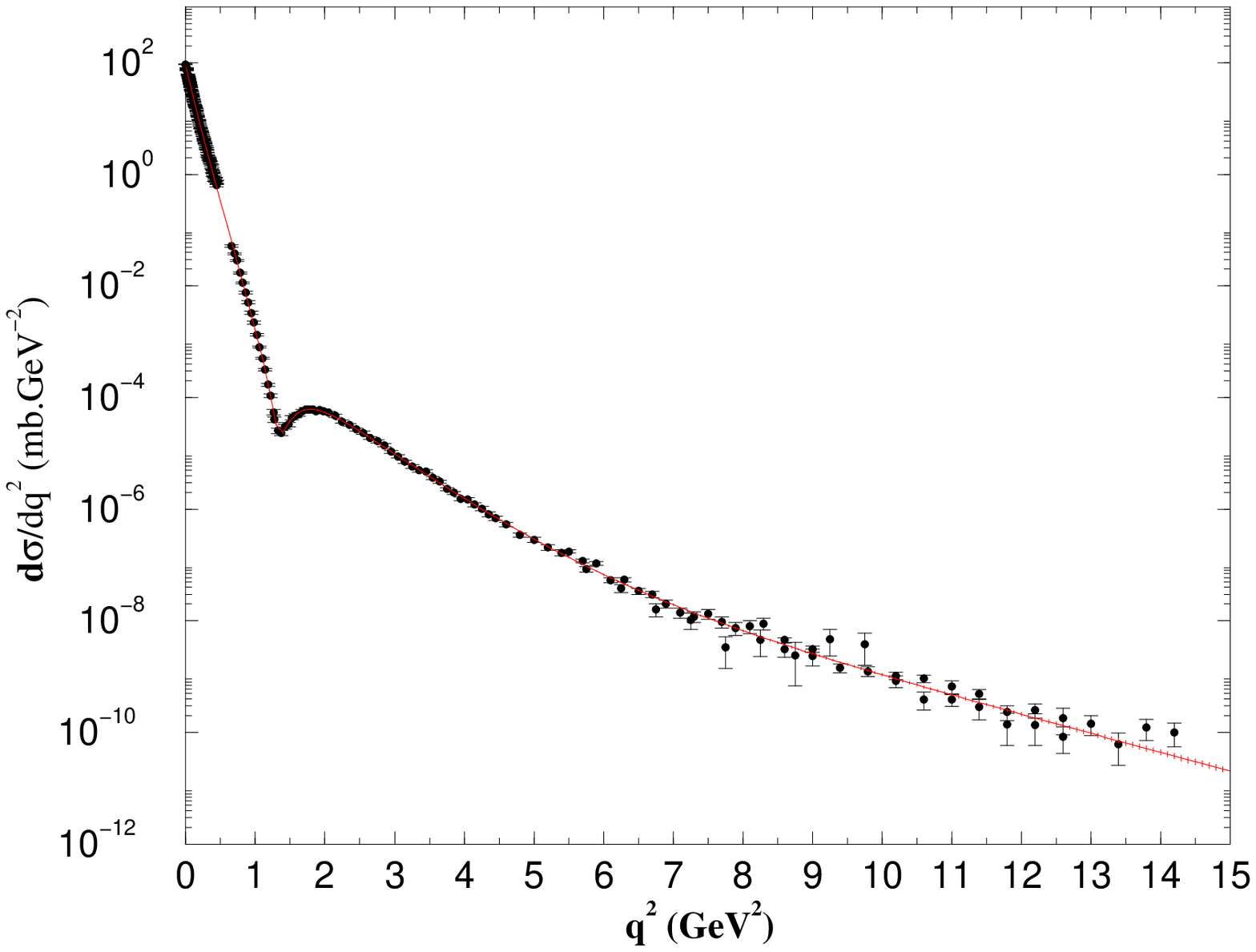,width=6cm,height=6.5cm}}
\vspace*{8pt}
\caption{Results of the fit to $pp$ differential cross section
data at $\sqrt s$ = 52.8 GeV (left) and adding data at 27.5 GeV (right).
Uncertainties regions are from error propagation.}
\end{figure}

\section{Proton Profile Function}

In the Impact Parameter Representation (Fraunhofer diffraction), the 
elastic scattering
amplitude at fixed energy is expressed by\cite{pred}
$
F(q) = \mathrm{i} \int_{0}^{\infty} bdb J_{0}(qb)
\ \Gamma(b)
$,
where $b$ is the impact parameter, $J_0$ is the zero order Bessel function 
(azimuthal symmetry assumed) and 
$\Gamma(b)$ is the Profile Function. By inverting the symmetrical Fourier 
transform we can determine  $\mathrm{Re}\ \Gamma(b)$ and
$\mathrm{Im}\ \Gamma(b)$ in terms of
$\mathrm{Im}\ F(q)$ and
$\mathrm{Re}\ F(q)$, respectively.
The results obtained by means of parametrization (3) and 
the fit up to $q_{max}^2 =$ 14 GeV$^2$ are shown in Fig. 2,
together with the uncertainty region determined by standard error 
propagation from the fit parameters.

\begin{figure}[th]
\centerline{\psfig{file=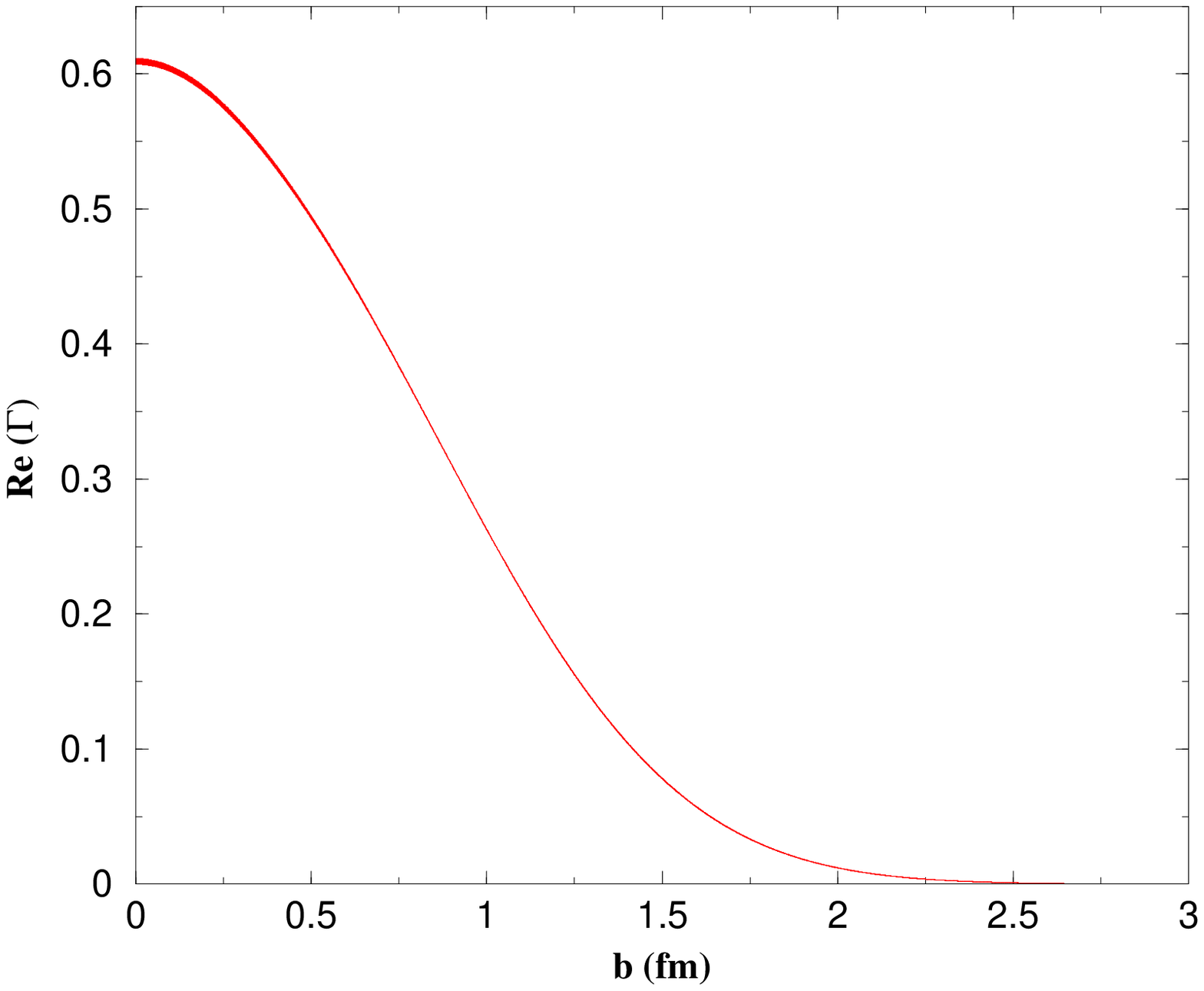,width=6cm,height=6.5cm}
\psfig{file=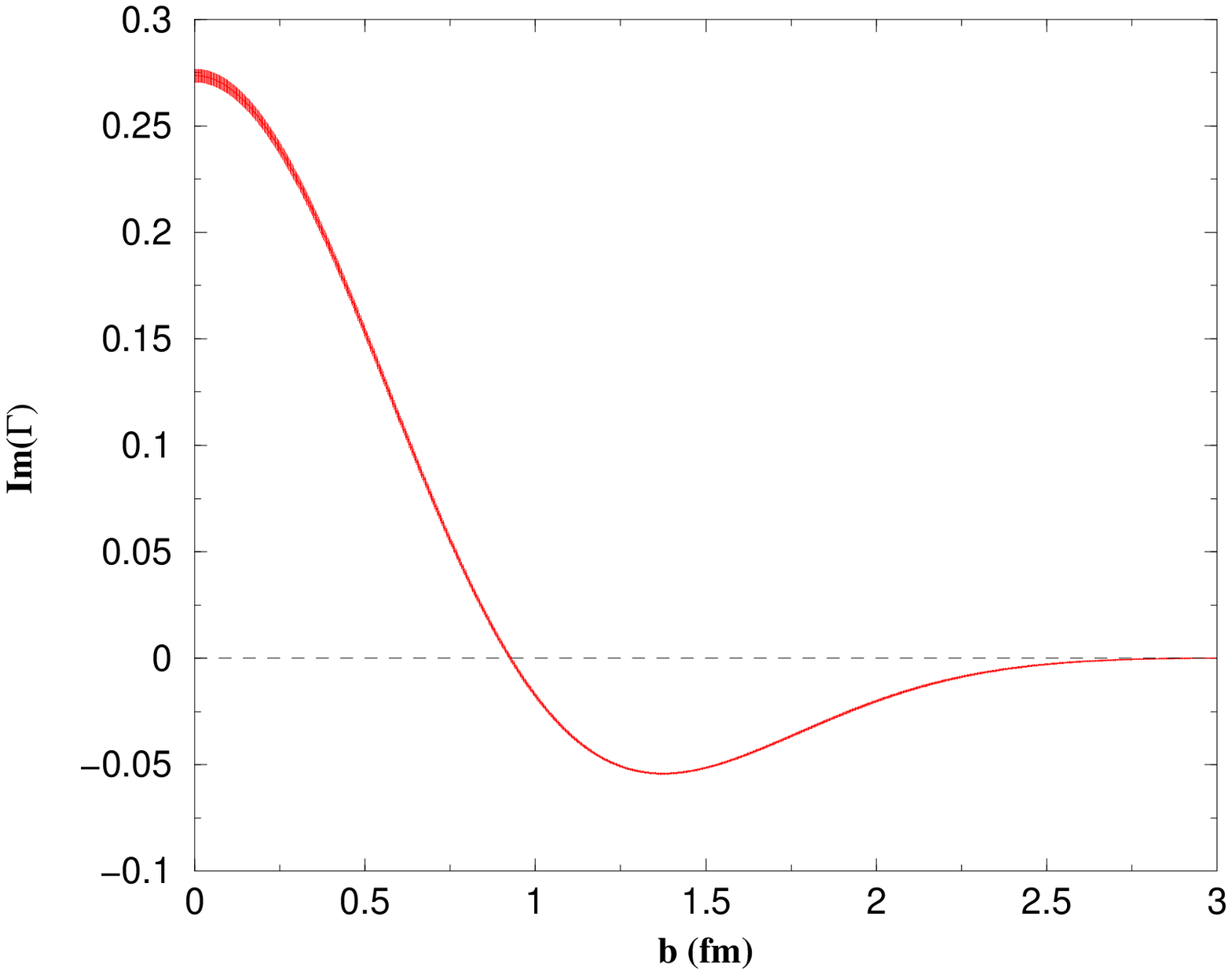,width=6cm,height=6.5cm}}
\vspace*{8pt}
\caption{Real and imaginary parts of the proton profile function
from fit to pp data at 52.8 GeV and 27.5 GeV (right panel in Fig. 1).}
\end{figure}

\section{Conclusions}

From Fig. 2 we see that, within the uncertainty regions, both the real and imaginary
parts of the profile present a change of curvature around 1 fm, with the
imaginary part becoming negative above this point (1 - 3 fm). We can interpret the result
as two different (dynamical) contributions, one associated with a dense
central region and another with an evanescent peripheral region. This
naive interpretation, however, seems not in
disagreement with the picture of a
hadron as a valence quark core surrounded by a gluon cloud.

\section*{Acknowledgments}

We are thankful to SAE - UNICAMP 
and Fapesp (Contracts No. 03/00228-0 and No. 04/10619-9)
for financial support.

\end{document}